\documentclass[12pt]{article}
\usepackage{times}
\usepackage[authoryear]{natbib}
\usepackage{graphics}
\begin{document}
{\LARGE\flushleft The potential of near-infrared high-resolution 
studies on\\[0.3ex]
the field of galaxies in the young universe}\\[2ex]
{Markus Wittkowski\\[0.2ex] European Southern Observatory, 
Casilla 19001,
Santiago 19, Chile, mwittkow@eso.org}\\[2ex]
September 30, 2001\\
\paragraph*{Abstract}
So far, high resolution techniques on the one hand provide morphological 
information on bright nearby objects. On the other hand, telescopes with 
large collecting areas allow us to detect very faint and distant objects, 
but not to obtain a spatial resolution which is sufficient for detailed 
morphological studies. Currently, the construction of large optical and 
infrared interferometers like the Keck Interferometer, the Very Large 
Telescope Interferometer (VLTI) and the Large Binocular Telescope 
Interferometer (LBTI) is in progress. These instruments will simultaneously 
provide larger collecting areas and higher spatial resolutions than current 
instruments. Thus, they might enable for the first time near-infrared studies 
of galaxies in the young universe with an absolute spatial resolution as 
available today only for the closest galaxies. 
Using recent results in the field of high resolution studies of nearby 
galaxies, a rough idea of what might be expected to be observed is given.
The concepts of the forthcoming interferometers is reviewed and
technical aspects that are essential for observations of distant galactic
centers are discussed. An outlook is given on which observational tasks can 
be addressed to these instruments and how the results will increase our 
knowledge in the field of the evolution of galaxies and the universe.
\section{Introduction}
\begin{minipage}{1.0\textwidth}
\hfill%
\begin{minipage}{0.85\textwidth}
''For all cosmic objects, we would ideally like to know the flux 
density over the entire electromagnetic spectrum. This goal is now almost 
achieved, at least for the brightest objects in each important class. The 
obvious next goal is to map out each object's structure in as much detail as 
possible.''\\[-0.3cm]
\begin{minipage}{1\textwidth}
\hfill%
Martin R. Rees (2000)
\end{minipage}
\end{minipage}
\end{minipage}\\[2ex]
As nicely expressed by Martin R. Rees (2000), high-resolution studies 
directly revealing the morphology of astrophysical objects are an 
essential tool to further our understanding of all parts of 
astrophysics, and have so far often been the key to milestone achievements.
While studies with highest resolutions down to the order of 1\,mas using 
interferometric techniques became already routine tools at radio wavelengths 
with facilities like the VLBI (Very Large Baseline Interferometer), large
optical and infrared interferometers are in the process of being 
constructed right now. Several optical and infrared interferometric 
instruments with relatively small collecting areas have been operated 
starting with the pioneer works by Fizeau (1868, 1873) and 
Michelson (1890)\footnote{An excellent overview on past and present optical 
long-baseline interferometry is provided by Peter Lawson at 
``olbin.jpl.nasa.gov''.}. The first 
construction of large interferometric facilities with 8-10\,m class 
telescopes and baselines up to the order of 100\,m is currently in progress 
with the Keck, VLTI (Very Large Telescope Interferometer), and LBT 
(Large Binocular Telescope) interferometers. First fringes at the Keck 
interferometer were achieved in February 2001 using test siderostats and 
in March 2001 using the two 10\,m telescopes (JPL News Release from 
March 14, 2001). First fringes at the VLTI using test siderostats followed 
closely in March 2001 (ESO Press Release 06/01 from March 18, 2001)
and the first beam combination using the 8\,m unit telescopes is 
envisioned for November 2001. 
The LBT will provide a complimentary interferometric facility with two 
8.4\,m telescopes on one mount from about 2004. Further interferometric 
beam combinations as those between the Subaru telescope and a neighbor have 
been discussed (Nishikawa et al. 1998). These facilities with their large 
collecting areas will be the first instruments that allow us to study galactic 
centers with spatial resolutions down to the order of 1\,mas at optical and 
infrared wavelengths. With the 120\,m 
VLTI baseline the resolution $\lambda/B$ at 1.2\,$\mu$m is 2\,mas, i.e. 
about the same as currently achieved at radio wavelengths with VLBI.
This angular resolution will allow us to study close galactic 
centers at optical and infrared wavelengths with there unprecedented absolute 
spatial resolution, as well as distant galactic centers with absolute spatial 
resolutions as available today for only the closest galaxies. The feasibility 
of the latter is discussed in this article. Previous publications mentioning
optical interferomtric observations of high-redshift objects
include Voit (1997), Tytler (1997), and Angel et al. (1998). 
In Sect.~\ref{sec:interf} general principles of interferometry are briefly 
reviewed and their implications on the feasibility of this project are 
discussed, followed by a review of the concepts of the actual interferometers 
and instruments. The center of the Seyfert galaxy NGC\,1068 is one of the 
best studied nearby active galactic nuclei (AGN). High-resolution studies 
of this object at the diffraction-limit of current single telescopes are 
reviewed in Sect.\ref{sec:diffz} in order to provide a first rough idea of the 
structures that can be expected to be observed in the case of high-redshift 
galactic centers.  This object is then put to different redshifts, and 
observational limits by apparent magnitude and spatial resolution are 
analyzed. Finally, in Sect.~\ref{sec:summ}, these results are summarized and 
additional issues are mentioned, as further scientific objectives and 
observational prospects in the more distant future. The latter include the 
use of extremely large telescopes (ELT) with diameters of the order 
of 100\,m. Several concept studies of such projects have been started,
for instance the 100\,m ''OWL'' (for {\it Overwhelmingly Large Telescope}) 
project (see Dierickx \& Gilmozzi 2000 for the conceptual design
and Gilmozzi 2000 for science opportunities).
\section{Interferometry}
\label{sec:interf}
Until extremely large optical telescopes with diameters of the order 
of 100\,m become available, optical and infrared studies with spatial 
resolution down to about 1\,mas require the use of interferometers. 
Interferometric measurements record the amplitude and phase of interference 
fringes rather than directly providing images. The basic implications of 
this technique on the feasibility of our project is discussed in the 
following. The characteristics of the different interferometers and their 
instruments is given at the end of this section. 
\paragraph{Field of view}
Two types of interferometry have to be distinguished, namely the
image plane or ''Fizeau'' and the pupil plane or ''Michelson'' 
beam combinations\footnote{The terms image plane and Fizeau beam combination,
as well as pupil plane and Michelson beam combination are sometimes not
used synonymously. In fact, it is in principle possible to achieve
an image plane beam combination with a Michelson style interferometer,
as mentioned later in this article}. 
The former technique implies that the output pupil
equals the input pupil or a scaled version of it for the whole integration
time, so that the beams can interfere at the final (output) image plane. 
This configuration can most easily be achieved if all apertures share one 
mount, and the entrance pupil is constant for all pointing directions.
Then, the field of view is limited only by the performance of the optical
components, e.g. the adaptive optics system. Such a configuration was chosen
in the case of the LBT interferometer, and, at near-infrared wavelengths,
a field of view of one arcmin or more is expected (Herbst et al. 2000). 
The pupil plane or ''Michelson'' beam combination technique, as chosen for
the Keck and VLT interferometers, a configuration where input and output
pupils are not equal, limits the field of view to about 1 arcsec.
It has the advantage of allowing long baselines. With this technique, 
spatial filtering is often applied in order to improve the accuracy of the 
measured visibility values, which further limits the field of view to the 
size of one airy disk. 
This field of view will usually be sufficient, since the interest 
concerns scales which are not available with single telescopes at their 
diffraction limit. Admittedly, for objects which are themselves not bright 
enough to allow for adaptive optics and fringe tracking, as those discussed 
in this article, a limited field of view decreases the probability to find
a bright reference star within this field. To cope with this, Michelson style 
interferometers can be equipped with a so called ``dual feed facility'' which 
allows us to feed two fields into the beam combiner. The second field can 
then be chosen to include the reference star. The maximum separation of the 
two fields is again limited by the optical quality and, for the example of 
the VLTI, amounts to about 1\,arcmin, which equals the field of view of Fizeau 
type interferometers.  
\paragraph*{Magnitude limit}
\begin{table}
\caption{Cross correlation of high-redshift ($z\ge0.5$) objects in the
catalog ''Quasars and Active Galactic Nuclei (8th Ed.)(Veron-Cetti \& 
Veron 1998)'' with objects in the HIPPARCOS catalog (Perryman \& ESO 1997) 
with a distance of up to 1\,arcmin. The table lists the name of 
the AGN, its redshift $z$, its apparent $V$ magnitude $V_\mathrm{AGN}$, its 
bolometric luminosity
M$_{\rm bol}$, the distance to the HIPPARCOS star $d$, the HIPPARCOS catalog
number HIP, the star's spectral type Sp.T., and its apparent $V$ magnitude 
$V_\mathrm{Ref}$.}
\vspace*{1ex}%
\begin{tabular}{llrrr|l|rlr}
AGN   &$z$   &   &$V_\mathrm{AGN}$  &M$_{\rm bol}$ & $d$ [arcsec] & 
HIP & Sp.T. & $V_\mathrm{Ref}$ \\ \hline
B19.09      &2.02&     &18.2&-27.8&58.9  &
9296       &G8III& 6.9 \\
PKS 0435-300&1.33&     &17.5&-27.4&29.2  &
21550      &F5V  & 8.1  \\
OK 568      &0.57&     &18.6&-24.1&28.3  &
47818        &K0   & 8.7  \\
AH 26       &2.21&     &18.6&-27.5&20.3  &
64269       &K0III& 8.4  \\
PKS 1551+130&1.29&     &17.7&-26.8&54.8  &
77830        &F0   &11.4  \\
\end{tabular}
\label{tab:refstars}
\end{table}
The incoming wavefront above each aperture is corrupted by atmospheric 
phase noise. Hence, each beam has to be corrected for this
effect using the technique of adaptive optics before the beams are
combined. This requires that the wavefront has to be analysed on
sub-apertures with the size of the Fried parameter $r_0$, 
limiting the magnitude 
of the reference star. Despite this correction, the optical path difference 
between two telescopes varies due to different mean phase values. Hence, 
the fringes have to be tracked, which limits the magnitude as well. Either 
the object itself or a nearby reference star within the isoplanatic patch
and the field of view has to be bright enough to allow for adaptive optics 
and fringe tracking. The critical value is about 12\,mag at 2.2\,$\mu$m for 
the planned interferometers. Since the objects in discussion are generally 
fainter than this, the use of a reference star for fringe tracking and 
adaptive optics will be mandatory. Table \ref{tab:refstars} lists all objects 
given in the catalog {\it Quasars and Active Galactic Nuclei (8th Ed.)}
(Veron-Cetti \& Veron 1998) with a redshift $z$ of larger than 0.5 that have 
a star given in the HIPPARCOS catalog (Perryman \& ESA 1997) within a distance 
of less than 1\,arcmin. Assuming that a field of view of 1\,arcmin
can be reached, or that a dual feed technique is available (see above), this 
tests whether such reference stars can be found for the objects in question. 
A total of 5 AGN/reference star pairs were found. Considering that the Veron 
catalog is not complete and that the HIPPARCOS catalog is complete only 
up to a $V$ magnitude of 9, while reference stars up to $K$ magnitude 12, 
i.e. $V > 12$, can be used, this shows that a sufficient number of AGN with 
such a reference star will be available.
\paragraph{Imaging capabilities}
Interferometers measure the amplitude and phase of interference 
fringes. In absence of atmospheric and instrumental phase noise, this is 
the amplitude and phase of the complex object visibility, i.e. the fourier 
transform of the object intensity distribution at a spatial frequency 
corresponding to the length and orientation of the used baseline. 
In presence of phase noise, however, an incoherent average of the measured 
complex visibility values would be caused to zero. 
The use of a fringe tracker can stabilize the fringe position, but not to an 
extent that the single phases could be used for imaging.  
A way to directly derive the fringe phase with sufficient accuracy is the 
method of ''phase referenced imaging''. This technique requires a dual feed 
facility in combination with a metrology system. Then, the differential delay 
between the fringe positions of a bright reference star and of the faint 
object is measured, whereas the bright star fringe phase serves as a reference 
for the object fringe phase. However, this technique is technically difficult 
to realize.
Generally, unbiased estimators exist for the squared visibility 
amplitudes and for the closure phases, the 
phase of the triple product or bispectrum, the product of three complex 
visibilities corresponding to baselines that form a triangle. 
The object visibility amplitude can than be 
derived as the square root of the squared visibility amplitude, despite the 
need of attention to the asymmetric error bars and observed squared visibility
values that may have negative values due to noise. In case that measurements 
were done at all spatial frequencies between zero and a maximum baseline, and 
with all orientations, the phases of the complex object visibility can be 
derived by recursion methods. This full ''$uv$-coverage'', can be achieved 
with a configuration like that of the LBT interferometer. 
Once all complex object visibility values have been recovered, the object 
intensity distribution can be obtained by fourier-re-transformation. 
If the $uv$-coverage is not complete, the 
object intensity distribution can be reconstructed using imaging techniques 
which effectively interpolate this limited $uv$-coverage, as applied in 
radio interferometry. 
The signal-to-noise ratio of the final image 
decreases with decreasing $uv$-coverage. It decreases as well with 
increasing number of resolution elements, i.e. with the complexity of the 
observed object. Hence, the imaging of 
complex source structures, as might be expected for extended galaxies 
might turn out to be impossible with interferometers that provide only a 
limited $uv$-coverage, as the VLTI or Keck interferometer. Here, imaging
or model fits that determine only a few basic parameters of the compact 
central cores might be a better choice. In the context of 
high-redshift galactic centers, such simple compact structures could be the 
expected dust tori, the inner parts of jets, or structures between broad line 
and narrow line region. A more detailed discussion of the structures that 
might be expected follows in section \ref{sec:ngc1068}.
\paragraph*{Available interferometers and their instruments}
\begin{table}
\caption{Overview on first generation interferometric instruments for the
Keck, VLT, and LBT interferometers. Given is the name of the interferometer 
with its maximum baseline; and the name of the instrument with the operating
wavelength, the limiting magnitude and the approximate date of first fringes.
For further information and references see the text.}  
\vspace*{1ex}%
\begin{tabular}{lr|lllr}
Interferometer & max. basel. & Instrument & $\lambda$ [$\mu$m]& lim. magn. & first fringes \\ \hline
     &     & two-way & 1.5-2.5 & & \\
Keck &  85\,m & multi-way & 1.5-5 & & \\
     &     & nulling & 10 & & \\\hline
     &     & AMBER & 1-2.5 & K $\sim$ 12/20 & 2003 \\
VLTI & 130\,m & MIDI  & 10-20 & N $\sim$ 3/8   & 2002 \\
     &     & PRIMA &                  &    & 2004 \\\hline
LBTI &  23\,m & LINC  & 1-2.4 & K $\sim$ 20-26 & 2004 \\ 
\end{tabular}
\label{tab:instr}
\end{table}
Interferometers in construction that will provide collecting areas large enough
to allow observations of galactic centers include the Keck, VLTI, and LBT
interferometers. The Keck interferometer is located on Mauna Kea, Hawaii,
the VLTI on Cerro Paranal, Chile, and the LBT on Mt. Graham, Arizona.
The Keck and VLT interferometers are ''Michelson'' style (see above) 
facilities and provide relatively large maximum baselines of 85\,m and 
130\,m, respectively. As a result, the available $uv$-coverage (see above) 
is not complete, and the field of view is limited to about 1-2\,arcsec. 
However, ''dual feed facilities'' (see above) are planned, that allow the use 
of a bright reference star for adaptive optics and fringe tracking. 
Planned Keck instruments include near-infrared two-way combiners for 
astrometry, two-element fringe tracking, and co-phasing; a near-infrared 
multi-way imaging combiner; and a mid-infrared nulling combiner 
(Colavita \& Wizinowich 2000).
First VLTI instruments include the mid-infrared instrument MIDI and 
the near-infrared instrument AMBER, followed by the dual feed facility PRIMA 
(Glindemann et al. 2000). AMBER will allow the combination of 
three beams, and thus will provide imaging capabilities. Using PRIMA to 
provide a bright (K $\sim$ 12) reference star, AMBER will have a 
limiting K magnitude of about 20, and MIDI a limiting N magnitude of about 8. 
PRIMA can also be used for phase referenced imaging (see above),
allowing imaging with both MIDI and AMBER, for fainter objects 
than with closure phase techniques (Delplancke et al. 2000).
The LBT interferometer is a ''Fizeau'' style facility with two 8.4\,m 
telescopes on one mount. This facility will provide a fairly 
complete $uv$-coverage and a relatively large field of view of 1\,arcmin,
on the cost of a more limited spatial resolution corresponding to a maximum
baseline of 23\,m. An interferometric beam combiner, LINC, is planned as
one of the first light instruments. It will operate at wavelengths from 
1\,$\mu$m to 2.4\,$\mu$m with a limiting K magnitude of 26.3 for a point 
source detection and 19.6 for the reconstruction of extended sources, under 
certain assumptions (Herbst et al. 2000).
Table \ref{tab:instr} gives on overview of the planned VLTI and LBTI
instruments of the first generation. While a nulling beam combiner,
which can exclude the light from the central point source, is planned
as a first generation instrument at the Keck interferometer, such an insturment
is in discussion as a second generation instrument for the VLTI.
Further discussions on future instruments include homothetic pupil mapping for 
Michelson style interferometers, fiber optics with coherent amplifiers to 
send the light to the beam combination lab, and artificial guide stars for 
fringe tracking.
\section{Expected structures and observational limits}
\paragraph*{The Seyfert Galaxy NGC\,1068 as a prototype galactic center}
\label{sec:ngc1068}
The Seyfert galaxy NGC\,1068 harbors one of the brightest and closest 
active galactic nuclei (AGN). Hence, it is one of the best studied galactic
centers and observations of this object, combined with theoretical
modeling like radiation transfer calculations, have substantially contributed 
to our current view on the structure of AGN. In fact, it is one of the objects 
studied in the classical work by Seyfert (1943) and it is the archetype of
type\,2 Seyfert galaxies. The observation by Antonucci \& Miller (1985) 
exhibiting broad emission lines in polarized light, initiated the unification 
scheme for AGN (see Antonucci 1993 for a recent review). In particular,
high-resolution studies of this object have been performed at a wide range
of wavelengths, testing and extending our view on the unification scheme. 
In this article, high-resolution information on NGC\,1068 is used to provide 
an idea on which structures might be expected to be observed in the case of 
more distant galactic centers with higher angular resolution, i.e. with 
the same absolute spatial resolution as available today for NGC\,1068. 
HST observations with an angular resolution of $\sim$\,0.1\,arcsec were 
obtained at the UV continuum, the optical continuum and the [OIII] 501\,nm 
emission line (Macchetto et al. 1994). These images show the narrow-line 
region, a conus-shaped structure, with the central source being obscured. 
Radio observations with similar angular resolution were performed by 
Ulvestad et al. (1987), Gallimore et al. (1996a,b), Muxlow et al. (1996). 
These observation show several radio peaks in approximately NE-SW direction, 
of which only one (''S2'') shows a positive spectral index 
$\alpha$ ($S_\nu\sim \nu^\alpha$), and is considered the position of the 
central core. The other emission peaks are connected with
the radio jet. The central source S2 was studied by Gallimore et al. (1997)
at 8.4\,GHz with a very high angular resolution of $\sim$\,1\,mas, revealing 
a structure approximately perpendicular to the radio jet, which is 
interpreted as emission from the torus surrounding the central nucleus. 
Mid-infrared observations were carried out by Chelli et al. (1987), Bock
et al. (1998), Marco \& Alloin (1999), Alloin et al. (2000), 
Bock et al. (2000), showing a 
compact central core, extended emission in approximately N-S direction on 
short scales and structures corresponding to the radio peaks on larger scales.
Marco \& Alloin (1999) report an additional 80\,pc E-W structure 
along the position angle of the 8.4\,GHz radio continuum image, interpreted as
the torus at mid-infrared wavelengths. Near-infrared studies by 
McCarthy et al. (1982), Thatte et al. (1997), Rouan et al. (1998), 
and Weinberger et al. (1999) confirm a very compact central NIR core and 
extended emission in N-S direction with a size of the order of 10\,pc. 
The central 2.2\,$\mu$m core was resolved by bispectrum speckle 
interferometry at the diffraction-limit of the Russian 6\,m telescope, 
with a FWHM size of $\sim$\,30\,mas or $\sim$\,2\,pc for an assumed 
Gaussian intensity distribution and a flux of 
$F_K^{30\,\mathrm{mas}}=520\,\mathrm{mJy}\pm 210\,\mathrm{mJy}$ 
(Wittkowski et al. 1998). With a model of an unresolved source contributing 
50\% to the flux of the 30\,mas component (the maximum percentage consisting 
with the obtained visibility function), this compact component had a size 
of $\sim$\,45\,mas instead of 30\,mas, i.e. still being a very compact source 
with a size of a few parsec. More recent speckle observations seem to 
indicate that this compact core is asymmetric with a position 
angle of $\sim\,20^\circ$, and an additional more extended structure 
in N-S direction out to $\sim$ 25\,pc (Wittkowski et al. 1999), 
consisting with other NIR observations mentioned above. 
An interpretation for the very compact 30-45\,mas structure could 
be a dust component, e.g. at the wall of the conus, very close to the 
central source, scattering and emitting near-infrared light; possibly 
with an additional component of direct light from the central
source. The more extended $\sim$\,200\,mas structure might be central light 
scattered at electrons and dust further out in the ionization cone.  
The interferometers as described above work at near-infrared and 
mid-infrared wavelengths, and might only be able to map very compact sources,
due to the signal-to noise ratio after image processing. Thus, these very 
compact near-infrared structures with sizes of 2-4\,pc and the slightly
more extended strucutres out to tens of parsec, connected to the torus, 
the region between NLR and BLR, or the inner part of the jet, are most 
promising structures to be 
observed with these interferometers at more distant galactic centers. These 
measurements could then initiate extensions of our view on the structures at 
the very centers of AGN as a function of the universe's age. In the following 
paragraphs, the observational limits are discussed, when observing these 
NGC\,1068-like structures at higher redshifts. It should be kept in mind, 
however, that this is only a rough idea of what might be expected to be 
observed and that actual structures might be different.
\paragraph*{Observation of a NGC\,1068--like object at different redshifts.}
\label{sec:diffz}
Here, limits for the study of structures like those
observed at near-infrared wavelengths for NGC\,1068, but for different
redshifts, are derived based on the spatial resolution and the flux.
\begin{figure}
\begin{minipage}{0.64\textwidth}
\hfill%
\resizebox{0.49\hsize}{!}{\includegraphics{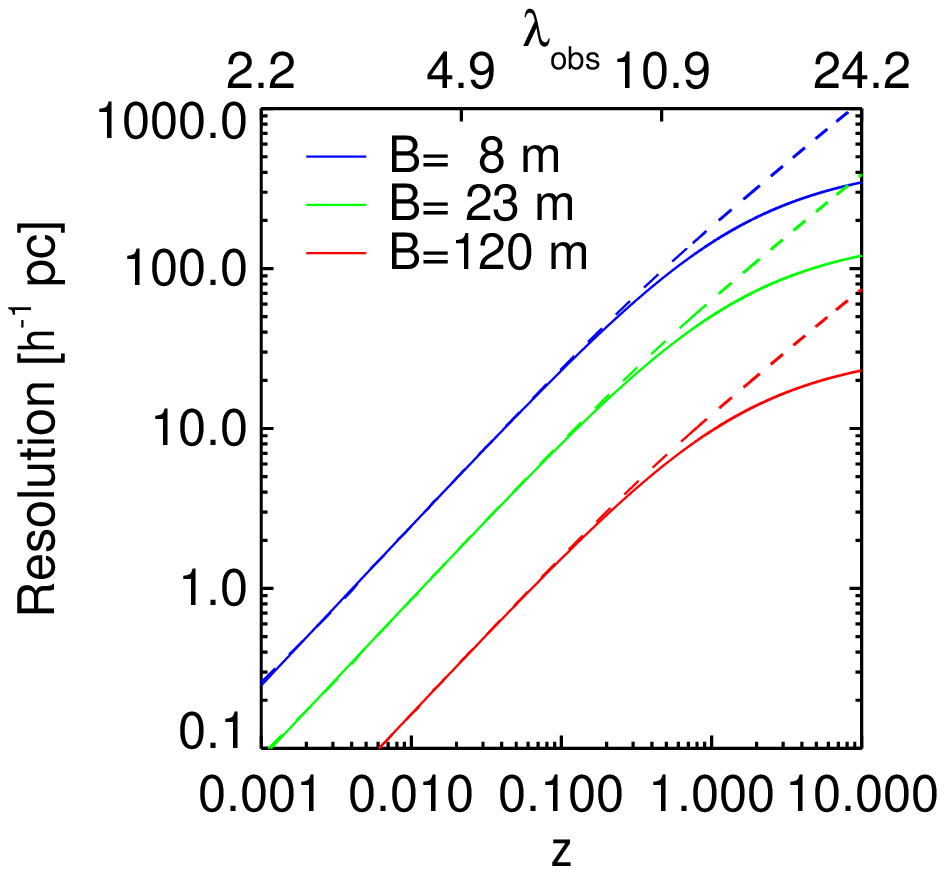}}
\hfill%
\resizebox{0.49\hsize}{!}{\includegraphics{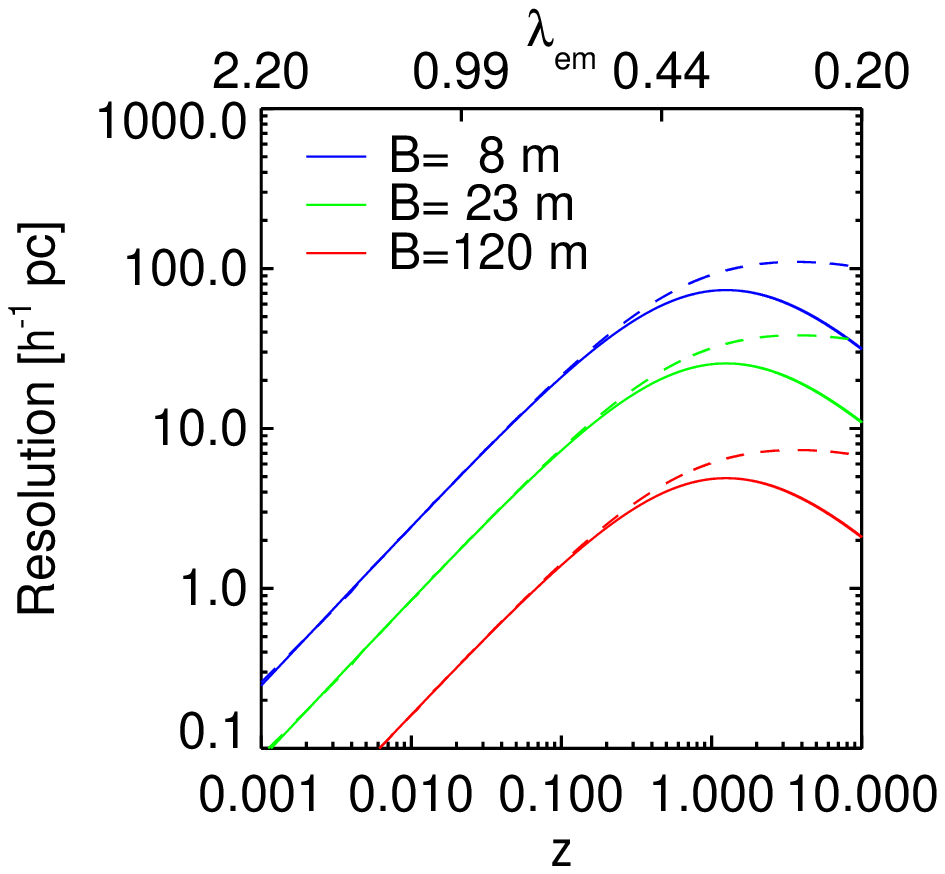}}
\hfill%
\end{minipage}
\begin{minipage}{0.35\textwidth}
\caption{Absolute spatial resolution based on Einstein-de-Sitter 
($\Omega_M$=1, solid line) and low density ($\Omega_M$=0.05, dashed line) 
world models. Left: Flux emitted at $\lambda_{\rm em}$=2.2\,$\mu$m; 
Right: Observing wavelength: $\lambda_{\rm obs}$=2.2\,$\mu$m. For the angular 
resolution, 0.3\,$\lambda/B$ was used, corresponding to a decrease of 
the squared visibility to 80\%.}
\end{minipage}
\label{fig:diffzres}
\end{figure}
\begin{figure}
\resizebox{0.32\hsize}{!}{\includegraphics{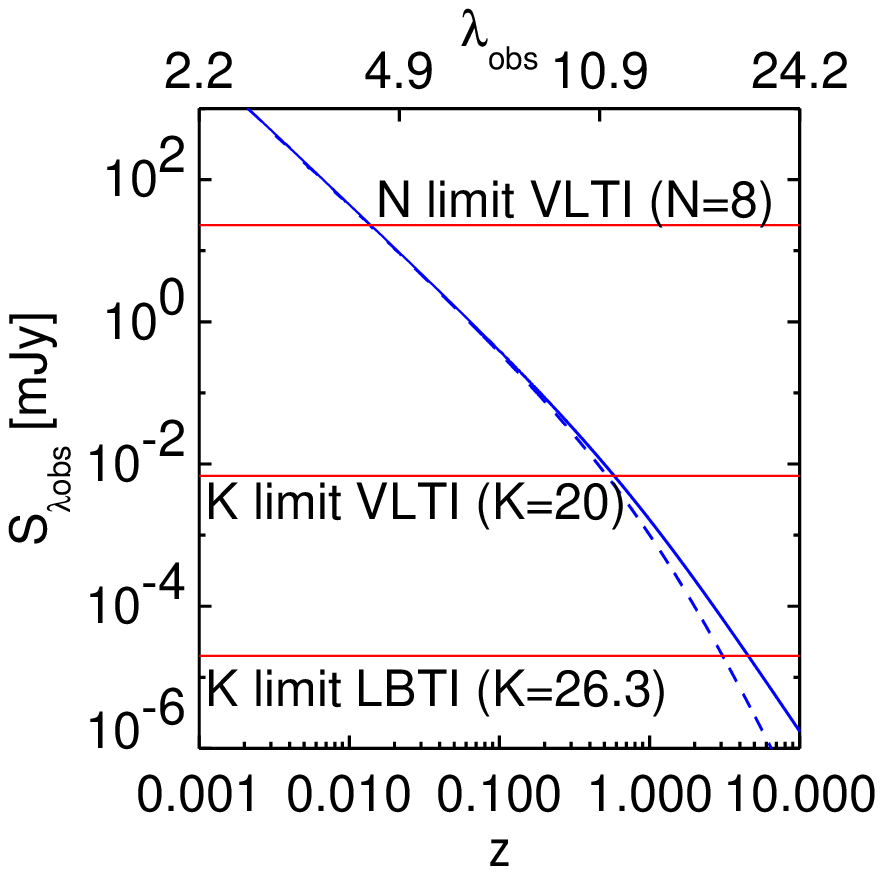}}
\hfill%
\resizebox{0.32\hsize}{!}{\includegraphics{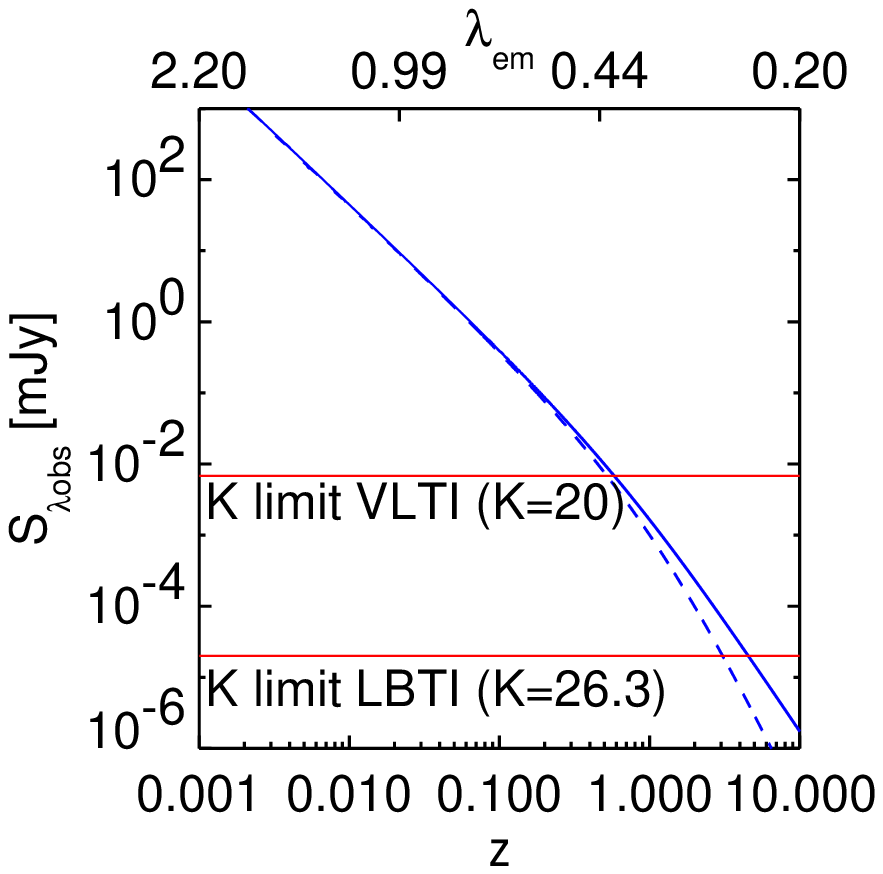}}
\hfill%
\resizebox{0.32\hsize}{!}{\includegraphics{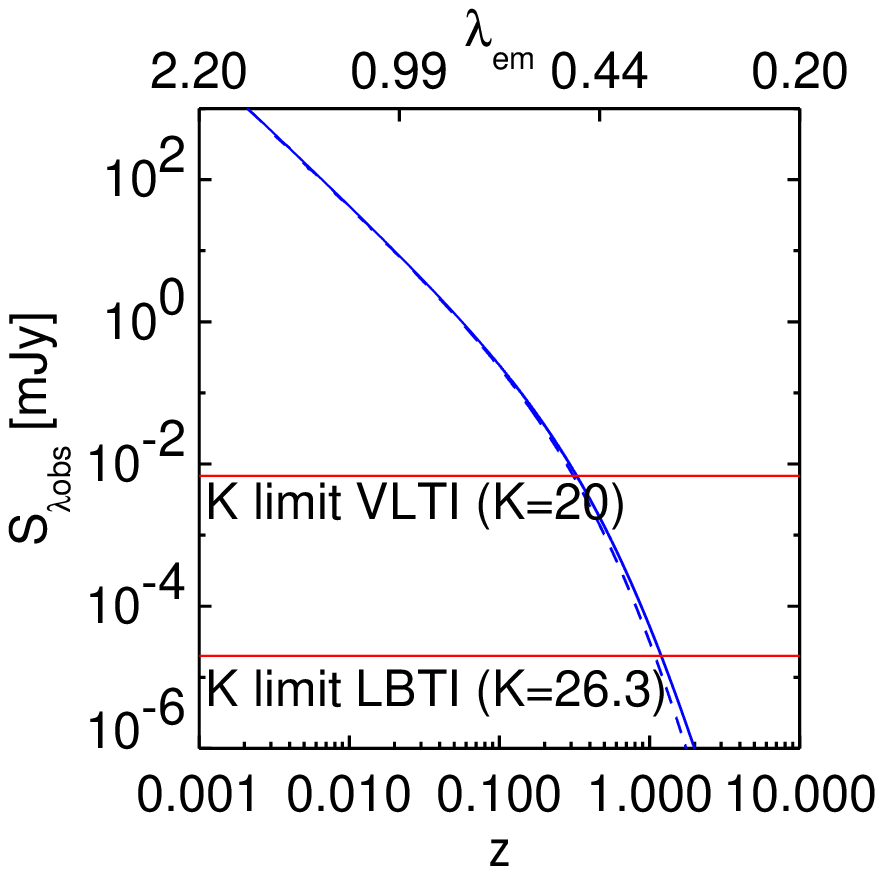}}
\caption{Flux of a source with $F_{\rm K}(z=0.003)=500$ mJy (compact
K-band component of NGC\,1068) as a function of redshift. Left: Flux is 
emitted at $\lambda_{\rm em}=2.2\,\mu$m; Middle and right: Observing 
wavelength is $\lambda_{\rm obs}=2.2\,\mu$, whereas (middle) is 
based on the assumption of a constant source spectrum, and (right) on that 
of a Planck source spectrum with $T=1500$\,K. 
As in Fig.~\ref{fig:diffzres}, the solid lines denote a Einstein-de-Sitter 
world model and the dashed lines a low density world model.}
\label{fig:diff2zres}
\end{figure}
\begin{table}[t]
\caption{Observational limits for the redshift $z$. Left: Limit by spatial 
resolution. Right: Limit by magnitude assuming a source flux 
of 0.5\,Jy at 2.2\,$\mu$m for $z=0.003$.}
\label{tab:limspres}
\begin{tabular}{lllll}
Structure   & 8\,m & 23\,m & 120\,m & for $z=1$\\ \hline
2\,pc, $\lambda_{\rm em}$ = 2.2\,$\mu$m   & 0.008 & 0.025 & 0.15 & 5\,km \\
20\,pc, $\lambda_{\rm em}$ = 2.2\,$\mu$m  & 0.08 & 0.25 & 2-6 & 90\,m\\[0.5cm]
2\,pc, $\lambda_{\rm obs}$ = 2.2\,$\mu$m  & 0.008 & 0.025 & 0.15 & 
3\,km\\
20\,pc, $\lambda_{\rm obs}$ = 2.2\,$\mu$m  & 0.1 & 0.4 & -- & 
50\,m \\
\end{tabular}
\hfill%
\begin{tabular}{lll}
    & Spectrum & Limit \\\hline
$\lambda_{\rm em}$  = 2.2\,$\mu$m &        & 0.05  \\
$\lambda_{\rm obs}$ = 2.2\,$\mu$m & const. & 0.6-4 \\
$\lambda_{\rm obs}$ = 2.2\,$\mu$m & 0.3-1 & Planck \\
\end{tabular}
\end{table}
\begin{table}[t]
\caption{Bolometric flux of NGC 1068 in comparison to all AGN in 
Veron \& Veron (1998) with M$_\mathrm{bol}$ $\le$ -29. It turns out that 
objects with highest bolometric luminosity are likely to be found at 
high redshifts.}
\label{tab:highmbol}
\begin{tabular}{lrrr}
Name & $z$ & $V$ & $M_\mathrm{bol}$ \\ \hline
NGC 1068 & 0.003 & 10.8 & -19.6 \\ \hline
Q 0042-2627 & 3.3 & 17.7 & -29.6 \\
DHM 0054-284 & 3.6 & 18.2 & -29.3  \\
Q 0130-403 & 3.0 & 17.0 & -30.0 \\
Q 0324-407 & 3.1 & 17.8 & -29.9  \\
\end{tabular}
\hspace*{1cm}%
\begin{tabular}{lrrr}
Name & $z$ & $V$ & $M_\mathrm{bol}$ \\ \hline
0846+51W1 & 1.9 & 15.7 & -29.4  \\
PG 1718+481 & 1.1 & 14.6 & -29.8  \\
PKS 2000-330 & 3.8 & 17.3 & -30.0  \\
Q 2204-408 & 3.2 & 17.6 & -29.1  \\
\end{tabular}
\end{table}
Figure \ref{fig:diffzres} shows the absolute spatial resolution that can be
achieved with baselines of 8\,m (single VLT UT telescopes), 23\,m (LBT), 
and 120\,m (VLTI) as a function of redshift. Shown are two cases, where (1)
light emitted at 2.2\,$\mu$m is observed, and (2) the object is observed
at 2.2\,$\mu$m. The former case regards to the near-infrared structures 
observed at NGC\,1068, which would be shifted to longer wavelengths. 
The latter case gives higher spatial resolution but it
is less clear whether there are structures feasible to observe since light
emitted at shorter wavelengths is studied, where the central source 
might be completely obscured. These graphs were obtained using the angular 
diameter distance given by Hogg (2000). Figure \ref{fig:diff2zres} shows 
analogously the observable 
flux of a source with 500\,mJy at 2.2\,$\mu$m for $z=0.003$ 
(compact NGC\,1068 structure) as a function of redshift, based on the 
luminosity distance given by Hogg (2000). The magnitude limits of the
interferometers discussed above are shown. For 2.2\,$\mu$m as the 
observing wavelength, an assumption on the source spectrum has to be made.
Shown are the results for a constant source spectrum -- a good
model for direct or scattered light originating from the central continuum 
source -- and for a Planck spectrum with T=1500\,K -- a model for
thermal radiation from dust components. Table \ref{tab:limspres} shows 
the resulting observational constraints by spatial resolution and magnitude.
It should be noted that the spatial resolution limit 
is a hard limit, whereas the magnitude limit might be more favorable since 
high-redshift objects are likely to have a higher intrinsic luminosity 
than NGC\,1068. Table \ref{tab:highmbol} shows that objects with large 
bolometric luminosity are likely to be found at high redshifts. 
Objects with a bolometric luminosity larger than that of NGC\,1068 
by a factor of $10^4$ can be found.
Observations of the compact 2\,pc structure are limited by spatial resolution 
to about $z=0.15$. The limit by magnitude based on the assumption of the
NGC\,1068 magnitude is about $z=0.05$ considering structures emitting 
at 2.2\,$\mu$m, and there is almost no flux limit for structures observed
at 2.2\,$\mu$m. In order to observe such a compact structure emitting at
2.2\,$\mu$m out to $z=1$, an interferometer consisting of 400\,m diameter 
mirrors and baselines of 3-5\,km would be needed.  
The observation of a 20\,pc NGC\,1068 structure in the young universe is 
not limited by spatial resolution, but only by magnitude, which
is $z=0.05$ for an emitting wavelengths of 2.2\,$\mu$m, and $z=1-4$ for
an observing wavelength of 2.2\,$\mu$m. Again, much brighter objects are
likely to be available, though.
\section{Summary and discussion}
\label{sec:summ}
The feasibility of observing galactic centers in the young universe at 
infrared wavelengths with very high angular resolution of up to 
about 1\,mas has been studied. Interferometers in construction that might 
allow such measurements for the first time have been discussed. Observations 
of the nearby AGN NGC\,1068 with the same absolute spatial resolution as
intended here for high-redshift objects, have been reviewed to give an idea
of the structures that might be expected to be observed. These structures
include a compact near-infrared core with a size of 2-4\,pc and a K-band
flux of 0.5\,Jy, and extended emission out to tens of parsecs. These 
components might be connected to the torus, the region between BLR and NLR,
and the inner part of the radio jet. Using these NGC\,1068 structures as a 
prototype, the observational limits by spatial resolution and magnitude 
for such structures at high-redshift were derived. Such measurements might aim
at testing and extending our unification scheme of galactic centers
including an evolution with the universe's age.
Results are given in Table \ref{tab:limspres}. However, the assumption
of NGC\,1068-like structures is considered only a starting point for these
studies, and, of course, structures might differ for high-redshift object.
For example, it was shown that high-redshift objects with much higher
intrinsic luminosity are likely to be found. This may also lead to different
length-scales of typical structures. Furthermore, there seem to be more
type\,I objects in the young universe than today, which might complicate
observations of structures which are relatively faint compared to less 
obscured central objects. There are also further scientific objectives
that have not been discussed in detail, but an estimate of their feasibility
might be supported by the discussions and figures given above as well. 
These objectives include observations of deep fields and of host galaxies 
using instruments with a fairly complete $uv$-coverage and large field of 
view like the LBT, with unprecedented spatial resolution. Another objective 
is the measurement of motions and parallaxes of galactic centers using 
phase reference 
techniques. If structures as discussed above are found, a further step in the 
more distant future could be the recording of spectra to derive velocity 
information and to determine black hole masses as a function of redshift.
Such scientific objectives might be supported by even more advanced
instrumental options in the more distant future. In the field of 
interferometry, these second-generation instrumentation might include
beam combinations with fiber optics and coherent amplifications to obtain
a better magnitude limit. A larger number of feasible objects could become
available by using artificial guide stars for adaptive optics and 
fringe tracking. An enlarged field of view for Michelson style
interferometers might be reached by homothetic mapping. Nulling 
interferometric instruments, which cancel out the light of the central 
source and image only off-center structures, might help to observe
faint structures close to not-obscured central sources. Polarimetric
interferometric observations might become available which would be of great
value since the origin of observed structures could be investigated in more
detail, i.e. whether structures are composed of reflected light, 
direct emission from the central source, or thermal emission.
Finally, observations
of these objects will become much easier with the construction of 
extremely large telescopes with diameters of up to 100\,m, like the OWL
telescope (see Sect.~1). In an even further step one might think of
interferometers with baselines on km scales consisting of OWL telescopes,
which was for instance mentioned by Glindemann et al. (2001) and called
the ''la OLA'' project (for {\it Overwhelmingly Large Array}). Such an
instrument would enable us to observe the compact NGC\,1068 2-4\,pc 
component, emitted at near-infrared and observed at mid-infrared wavelengths, 
out to a redshift of $z=1$.
\paragraph*{Acknowledgements}
It thank Sara Ellison, Andreas Glindemann and Markus Sch\"oller for 
information, help, and discussions during the preparation of this talk. 
I acknowledge previous discussions on related aspects with members of the 
''MPIfR Infrared Interferometry'' and NPOI groups.
\small
\setlength{\bibsep}{0.0ex}
\bibliographystyle{plainnat}

\end{document}